\documentclass[doublecol]{epl2}
\usepackage{amsmath}
\usepackage{amssymb}
\usepackage{color}
\usepackage{graphicx,subfigure}% Include figure files
\usepackage{dcolumn}% Align table columns on decimal point
\usepackage{bm}% bold math
\def\comment#1{}

\def\slashchar#1{\setbox0=\hbox{$#1$}           % set a box for #1 
   \dimen0=\wd0                                 % and get its size
   \setbox1=\hbox{/} \dimen1=\wd1               % get size of /
   \ifdim\dimen0>\dimen1                        % #1 is bigger
      \rlap{\hbox to \dimen0{\hfil/\hfil}}      % so center / in box
      #1                                        % and print #1
   \else                                        % / is bigger
      \rlap{\hbox to \dimen1{\hfil$#1$\hfil}}   % so center #1
      /                                         % and print /
   \fi}                                         %

\def\sigmab{{\mbox{\boldmath $\sigma$}}}

\def\sigmab{{\mbox{\boldmath $\sigma$}}}

\title{Deconfined Quantum Criticality and Conformal Phase Transition   
in Two-Dimensional Antiferromagnets
}
\shorttitle{Deconfined Quantum Criticality and Conformal Phase Transition}
 \author{Flavio S. Nogueira\inst{1} \and Asle Sudb{\o}\inst{2}}
 \shortauthor{F. S. Nogueira and A. Sudb{o}}
\institute{\inst{1} Institut f{\"u}r Theoretische Physik III, Ruhr-Universit\"at Bochum,
Universit\"atsstra\ss e 150, 44801 Bochum, Germany\\
\inst{2} Department of Physics, Norwegian University of
Science and Technology, N-7491 Trondheim, Norway}


\abstract{
Deconfined quantum criticality of two-dimensional $SU(2)$ quantum antiferromagnets featuring a transition from an antiferromagnetically ordered ground 
state to a so-called valence-bond solid state, is governed by a non-compact CP$^1$ model with a Maxwell term in 2+1 spacetime dimensions. We introduce a 
new perspective on deconfined quantum criticality within a field-theoretic framework based on an expansion in powers of $\epsilon=4-d$ for fixed number 
$N$ of complex matter fields. We show that in the allegedly weak first-order transition regime, a so-called conformal phase transition leads to a genuine 
deconfined quantum critical point. In such a transition, the gap vanishes when the critical point is approached from above and diverges when it is approached 
from below. We also find that the spin stiffness has a universal jump at the critical point.}

\begin{document}

\maketitle

\begin{figure}
\begin{center}
 \includegraphics[width=8cm]{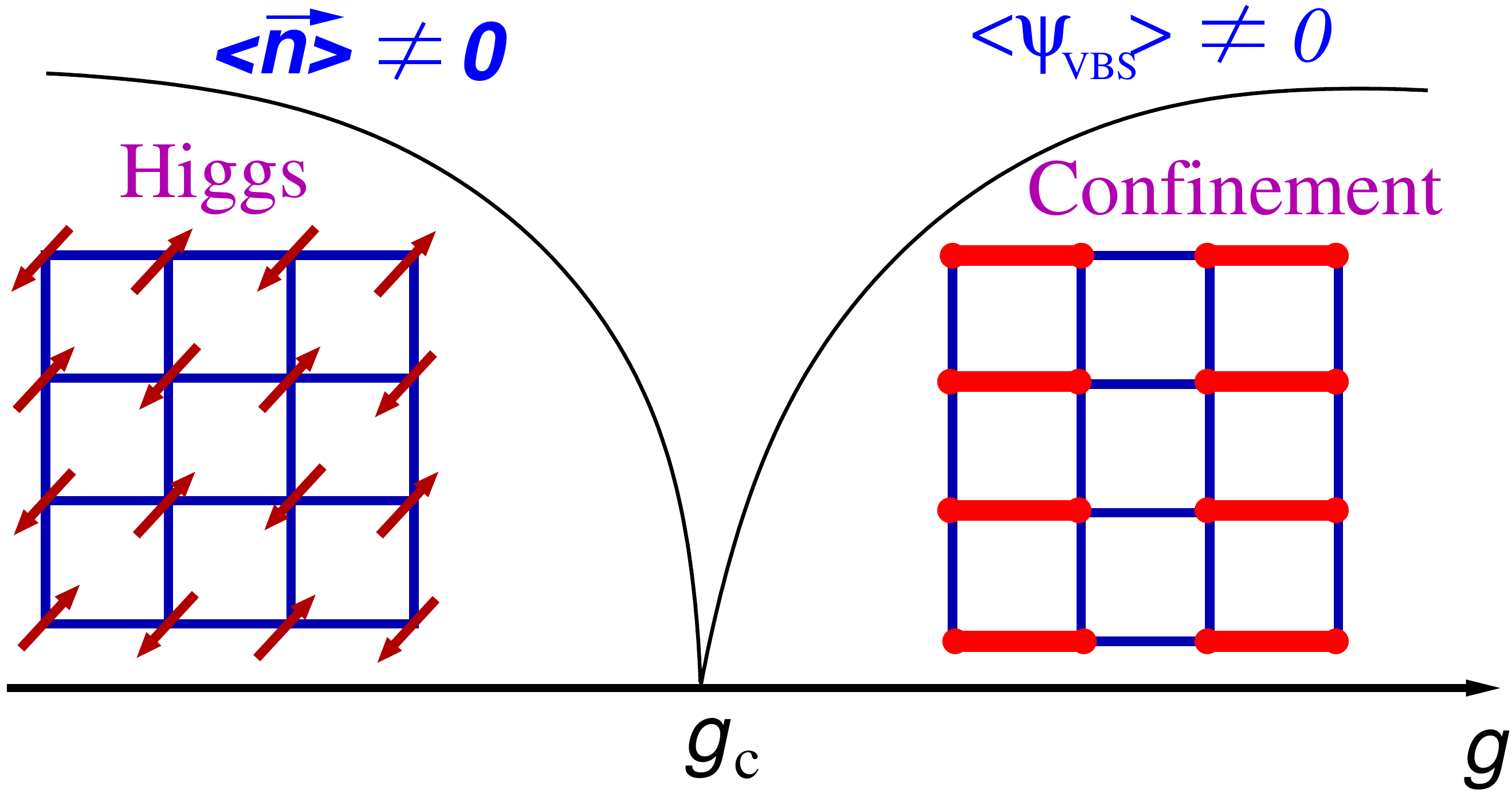}
\end{center} 
\caption{Schematic phase diagram for the $J-Q$ model [Eq. (\ref{Eq:J-Q})] showing a quantum phase transition between a N\'eel state and a VBS 
as a function of the dimensionless coupling $g=Q/J$. Both the N\'eel and the VBS order parameters are composed of spinon fields. On the lattice, 
they correspond to the composite fields ${\bf n}_i=(-1)^iz_{i\alpha}^*\sigmab_{\alpha\beta}z_i$ and to 
$\psi_{{\rm VBS},i}=(-1)^iL^{-1}\sum_jz_{i\alpha}^*z_{j\alpha}z_{j\beta}^*z_{i\beta}$, where $L$ is the number of lattice sites. In terms of the 
spinons $z_{i\sigma}$ both fields represent $U(1)$ gauge-invariant operators. In the Higgs phase, the spinons condense due to a spontaneous $U(1)$ 
symmetry breaking, leading to a N\'eel state.  In the confinement phase, all excitations are gapped and the spinons are confined, leading to a 
VBS state. In the N\'eel phase the emergent photon is gapped, while in the confined phase it is the dual of the emergent photon of the Higgs phase 
which is gapped.}
\label{Fig:AF-VBS}
\end{figure}

Many years have passed since a new paradigm for quantum phase transitions, the so-called deconfined quantum criticality (DQC) scenario, 
was introduced \cite{Senthil-2004}. In this paradigm, the effective quantum field theory does not contain any {\it elementary} fields 
representing the order parameters associated with the  underlying competing orders. It posits that in certain quantum 
phase transitions these order parameters are not elementary,  but composed of more elementary fields in the same way that 
in elementary particle physics mesons are constituted by quarks. The precise context where this happens involves competing orders featuring 
broken {\it internal} and spacetime symmetries. This occurs, for example, in certain $SU(2)$ quantum antiferromagnets (AF) where $SU(2)$-invariant 
interactions compete. A paradigmatic example is the so-called $J-Q$ model \cite{Sandvik_2007},
\begin{equation}
\label{Eq:J-Q}
 H=J\sum_{\langle i,j\rangle}{\bf S}_i\cdot{\bf S}_j -Q\sum_{\langle ijkl\rangle}\left({\bf S}_i\cdot{\bf S}_j-\frac{1}{4}\right)
\left({\bf S}_k\cdot{\bf S}_l-\frac{1}{4}\right),
\end{equation}
where both $J$ and $Q$ are positive. Defining the dimensionless coupling $g=Q/J$, we obtain the schematic phase diagram shown 
in Fig. \ref{Fig:AF-VBS}. For $g\ll 1$ the first term in (\ref{Eq:J-Q}) dominates, favoring a N\'eel state. For  
$g\gg 1$ the plaquette term in Eq. (\ref{Eq:J-Q}) dominates, favoring a valence-bond solid (VBS) ordered state. The N\'eel state breaks 
an internal symmetry, namely $SU(2)$. The VBS state preserves the $SU(2)$ symmetry while breaking the symmetries of the square 
lattice. As one broken symmetry is internal [the $SU(2)$ one] and the other one is spatial, quantum mechanics forbids 
their coexistence, since the VBS state is a long-range entangled state while the N\'eel state is long-range ordered. 

In the DQC scenario the operators measuring both N\'eel  and VBS order are comprised of more fundamental 
objects. These are the spinons, which are represented by an $SU(2)$ doublet of complex fields ${\bf z}=(z_1,z_2)$ satisfying the 
constraint $|z_1|^2+|z_2|^2=1$ at each lattice point. In terms of the spinon fields, the fields representing the N\'eel and VBS 
order parameters are $U(1)$ gauge-invariant objects. The gauge field arising in such a theory is an emergent ``photon''   originally 
defined on the lattice, and hence it is necessarily compact. This leads to instanton excitations that gap the dual photon  
(defined as $B_\mu=\epsilon_{\mu\nu\lambda}\partial_\nu A_\lambda$)  in the phase where the expectation value of the Higgs field 
is zero, i.e., the paramagnetic phase. This gap also corresponds to the mass of the instantons \cite{Polyakov}.  Thus, the 
VBS phase is one where the spinons are confined (see the text in the caption of Fig. \ref{Fig:AF-VBS}). In the N\'eel phase, 
on the other hand, the photon is gapped due to the Higgs mechanism.  One fundamental prediction of the DQC scenario is that the 
instanton-mass vanishes continuously for $g$ approaching a quantum critical point $g_c$ from above, thus suppressing them at 
the quantum critical point \cite{Senthil-2004}. For a version of this theory with easy-plane anisotropy \cite{Motrunich}, the 
suppression of instantons has been confirmed by Monte Carlo (MC) simulations \cite{Kragset}. In the easy-plane case, the suppression 
occurs in a weak first-order phase transition, and no quantum criticality ensues \cite{Kuklov_2006,Kragset}.   

For the $SU(2)$ DQC model, early MC results indicated a weak first-order phase transition \cite{Kuklov_2008}. Simulations performed on 
the $J-Q$ model have mostly yielded a second-order phase transition and signs of an emergent $U(1)$ symmetry 
\cite{Sandvik_2007,Sandvik-2010,Melko-Kaul_2008}, although a weak first-order phase transition has also been reported 
\cite{Jiang_2008}. Since the $J-Q$ model is one of the emblematic lattice models for the DQC scenario, a recent 
MC study \cite{Chen} made a comparative analysis of its phase diagram with the one obtained from the non-compact CP$^1$ model. 
While both models agree over a substantial portion of the phase diagram for moderate system sizes, they  behave differently 
at larger system sizes \cite{Chen}. Furthermore, there are indications that none of the models become critical, which would 
corroborate a weak first-order phase transition scenario. Recent large scale simulations \cite{Asle-NCCP1} on the non-compact 
Abelian Higgs model with CP$^1$ constraint indicate that the existence of a tricritical point cannot be ruled out. It is also 
worth mentioning that a large $N$-like MC study of the $J-Q$ and $J_1-J_2$ ($J_1$ nearest neighbor and $J_2$ next-nearest neighbor 
exchanges) models has been made recently \cite{Kaul-Sandvik}, aiming to compare with large $N$ limit of the CP$^{N-1}$ model, where 
quantum criticality is known to occur. In this study strong evidence for DQC has been found for $N>4$.  

This brings us to the main topic of this paper, namely, a quantum field-theoretic analysis of the non-compact Abelian Higgs model 
with a global $SU(N)$ symmetry. In the present context, there are two relevant versions of this theory, a non-linear and a linear 
one. The non-linear theory corresponds to a CP$^{N-1}$ model with a Maxwell term \cite{Lawrie_1983}, 
\begin{equation}
 \label{Eq:CP1(N-1)}
 {\cal L}_{{\rm CP}^{N-1}}=\frac{\Lambda^{d-2}}{\hat g}\sum_{\alpha=1}^N|(\partial_\mu-iA_\mu)z_\alpha|^2+\frac{1}{4e^2}F_{\mu\nu}^2,
\end{equation}
where $F_{\mu\nu}=\partial_\mu A_\nu-\partial_\nu A_\mu$ and $\sum_{\alpha=1}^N|z_\alpha|^2=1$. The 
linear version softens this constraint and has the more standard Higgs model form \cite{Senthil-2004}
\begin{eqnarray}
 \label{Eq:Higgs}
 {\cal L}_{\rm Higgs}&=&\sum_{\alpha=1}^N\left[|(\partial_\mu-iA_\mu)z_\alpha|^2+r|z_\alpha|^2\right]
\nonumber\\
&+&\frac{u}{2}
\left(\sum_{\alpha=1}^{N}|z_\alpha|^2\right)^2+\frac{1}{4e^2}F_{\mu\nu}^2.
\end{eqnarray}
Both models have the same symmetries. In parameter regimes where a critical point exists, they should belong to the same universality class. 
In the limit $e^2\to\infty$, both ${\cal L}_{{\rm CP}^{N-1}}$ and ${\cal L}_{\rm Higgs}$ have exactly the same critical behavior for 
large $N$  \cite{Hikami}. However, a recent calculation of the spin stiffness at large $N$ and finite  $e^2$  \cite{Nogueira-Sudbo-2012} showed 
that $\rho_s$ exponentiates to a Josephson scaling form only when $e^2\to 0$ or $e^2\to\infty$, corresponding to  $O(2N)$ or CP$^{N-1}$ universality 
classes, respectively. For finite values of $e^2$, the spin stiffness exhibits logarithmic violations of scaling \cite{Nogueira-Sudbo-2012},  
which have been reported in recent MC simulations of the $J-Q$ model \cite{Sandvik-2010,Damle}. 

Here, we address the actual nature of the phase transition in the gauge theory proposed to underpin deconfined quantum 
critical points. In gauge theories, Elitzur's theorem \cite{Elitzur} forbids the spontaneous breaking of a local gauge symmetry in any dimension. 
Therefore, there is no local order parameter available to distinguish phases. In MC simulations of the lattice version of model (\ref{Eq:CP1(N-1)}), 
one of the quantities studied is the spin stiffness \cite{Kuklov_2008}, which provides a nonlocal order parameter. However, a jump 
in the spin stiffness does not necessarily imply a first-order phase transition in this case. One could have a situation where the gap 
vanishes continuously as the critical point is approached, while the spin stiffness vanishes discontinuously. Some theories behave precisely in 
this way, a prominent example being the Berezinskii-Kosterlitz-Thouless (BKT) transition \cite{BKT}, where the inverse correlation length features an 
essential singularity at the critical point and no local order parameter exists \cite{MW}. In the case of the BKT transition, the superfluid stiffness 
has a {\it universal jump} at the critical point \cite{NelKost}. Theories with this type of behavior are said to undergo a 
conformal phase transition (CPT) \cite{ConfPhTrans}. Recent lattice simulations \cite{Deuzeman} show evidence of a CPT in $SU(N)$ gauge theories 
in $d=3+1$.  

We provide arguments to support a CPT scenario in DQC gauge field theories. First, we show that the $\epsilon$-expansion for the model (\ref{Eq:Higgs}) 
contains a regime where the inverse correlation length has an essential singularity and show that the spin stiffness features a universal jump at the 
critical point. Then, we derive a similar behavior for the mass of instantons in the paramagnetic phase of the CP$^{N-1}$ model (\ref{Eq:CP1(N-1)}).  

The one-loop RG $\beta$ functions are well known and were originally obtained using the Wilson RG \cite{Halperin-Lubensky-Ma}.  In the classic RG-analysis 
of Eq. \ref{Eq:Higgs} carried out in Ref. \cite{Halperin-Lubensky-Ma}, it was concluded that no stable infrared fixed point existed unless $N$ exceeded some 
large value $ N_c \approx 185$. For $N < N_c$, runaway flows of the RG-equations were found, and this was originally interpreted as a signature of a first-order 
phase transition. A more modern interpretation of the same, is that it signals the existence of a strong-coupling fixed point, and it is this point of view we 
take. Contrary to the scope of Ref. \cite{Halperin-Lubensky-Ma}, in this paper we undertake a careful analysis of the precise character of this strong-coupling 
fixed point. This has, to our knowledge, not been carried out. Such an analysis is of paramount importance, given the proposed DQC-scenario.  

To analyze Eqs. \ref{Eq:CP1(N-1)} and \ref{Eq:Higgs}, it is convenient to use the field theory RG with dimensional regularization in the minimal 
subtraction scheme \cite{ZJ}, rather than the Wilson RG  approach to the problem \cite{Halperin-Lubensky-Ma}. We introduce the renormalized dimensionless 
couplings  $f=m^{-\epsilon}e^2_R/(8\pi^2)$ and $g=m^{-\epsilon}u_R/(8\pi^2)$, where $\epsilon=4-d$ and $m$ is the Higgs mass scale related to the inverse 
correlation length. Here, $e_R^2$ and $u_R$ are the renormalized counterparts of the bare couplings $e^2$ and $u$. The asymptotic behavior of the renormalized 
gauge coupling will be crucial. In order to obtain it at one-loop order, we have to compute the vacuum polarization, $\Pi_{\mu\nu}(p)$, which 
yields the lowest order fluctuation correction to the Maxwell term in the action. In a $d$-dimensional spacetime, we have,
\begin{eqnarray}
 \label{Eq:eff-Maxwell}
 S_{\rm Maxwell}&=&\frac{1}{4e^2}\int d^d x F_{\mu\nu}^2
 \nonumber\\
 &+&\frac{1}{2}\int\frac{d^dp}{(2\pi)^d}
 \Pi_{\mu\nu}(p)A_\mu(p)A_\nu(-p),
\end{eqnarray}
where the vacuum polarization is obtained 
%from the Feynman diagrams in Fig. \ref{Fig:vacpol} 
as
\begin{eqnarray}
 \Pi_{\mu\nu}(p)&=&2N\delta_{\mu\nu}\int\frac{d^dk}{(2\pi)^d}\frac{1}{k^2+m^2}\nonumber\\
 &-&N\int\frac{d^dk}{(2\pi)^d}\frac{(2k-p)_\mu(2k-p)_\nu}{[(k-p)^2+m^2](k^2+m^2)}.
\end{eqnarray}
If dimensional regularization is used, the gauge symmetry is preserved along with current conservation, and therefore 
$\Pi_{\mu\nu}(p)$ is transverse. Thus,  $\Pi_{\mu\nu}(p)=\Pi(p)(p^2\delta_{\mu\nu}-p_\mu p_\nu)$. In the low-energy regime 
where $|p|\ll m$, we can evaluate all integrals explicitly for arbitrary $d$ to obtain 
$\Pi(p)\approx Nm^{d-4}\Gamma(2-d/2)/[3(4\pi)^{d/2}]$. The effective Maxwell contribution to the Higgs Lagrangian then becomes,
\begin{eqnarray}
 \label{Eq:L-Maxwell}
 {\cal L}_{\rm Maxwell}&\approx&\frac{1}{4e^2}\left[1+\frac{Ne^2}{3(4\pi)^{d/2}}\Gamma\left(2-\frac{d}{2}\right)m^{d-4}\right]F_{\mu\nu}^2.
\nonumber\\
& \equiv &\frac{1}{4e_R^2}F_{\mu\nu}^2,
 \end{eqnarray}
which defines the renormalized gauge coupling $e_R^2$. 
Recall that for $N=2$ and $e^2\to\infty$ the CP$^{N-1}$ model (\ref{Eq:CP1(N-1)}) is equivalent to the $O(3)$ nonlinear $\sigma$ model 
(see, for instance, Sect. 7.9 in Ref. \cite{Fradkin-book}), 
which has a second order phase transition. In fact, the CP$^{N-1}$ model exhibits a second order phase transition  
for any $N$ if $e^2\to\infty$. The same is expected to be true for the abelian Higgs model (\ref{Eq:Higgs}). Thus, let us consider 
the limit $e^2\to\infty$ of Eq. (\ref{Eq:L-Maxwell}) and its approximate form for small $\epsilon=4-d$,
\begin{eqnarray}
\label{Eq:L-Maxwell-1}
 {\cal L}_{\rm Maxwell}&\approx&\frac{N}{12(4\pi)^{d/2}}\Gamma\left(2-\frac{d}{2}\right)m^{d-4}F_{\mu\nu}^2
 \nonumber\\
 &\underset{{\rm small}~\epsilon}{\approx}&\frac{1}{4(8\pi^2f_* m^{\epsilon})}F_{\mu\nu}^2,
\end{eqnarray}
where $f_*=3\epsilon/N$. After approximating Eq. (\ref{Eq:L-Maxwell}) for small $\epsilon$, we obtain the one-loop  
$\beta$ function for the dimensionless gauge coupling $f$,
\begin{equation}
\label{Eq:beta-f}
\beta_f\equiv m\frac{df}{dm}=-\epsilon f+\frac{N}{3}f^2,
\end{equation}
and we identify $f_*$ as the infrared stable fixed point. Hence, for 
$2<d<4$ we have that $f$ approaches $f_*$ in two ways, namely, as 
$e^2\to\infty$ for fixed $m$ or as $m\to 0$ for fixed $e^2$. Therefore, we can use $(e^2)^{1/(4-d)}$ as the ultraviolet cutoff $\Lambda$. 

Note that $\beta_f$ is only a function of $f$. A two-loop calculation does not change this \cite{Folk}, and $\beta_f$ remains dependent only on $f$. Within 
dimensional regularization in the minimal subtraction scheme one may show that this holds to all orders, since the 
only poles in $\epsilon$ arise in diagrams containing uniquely powers of the gauge coupling. All other diagrams are finite 
for $\epsilon \to 0$. This follows from gauge invariance and Ward identities of the theory.   

The $\beta$ function for the coupling $g$ is given at one-loop order by \cite{Hikami} 
\begin{equation}
\label{Eq:beta-g}
\beta_g\equiv m\frac{dg}{dm}=-\epsilon g-6fg+(N+4)g^2+6f^2.
\end{equation}
There are two relevant regimes where critical points arise, depending on the value of the gauge coupling fixed point. 
For $f=0$, we have a nontrivial fixed point $g_*=\epsilon/(N+4)$ governing the critical behavior corresponding to the 
$O(2N)$ universality class, while the line $f=f_*=3\epsilon/N$ contains  a critical ($g_+$) and a tricritical ($g_-$) 
fixed point for $N>N_c=6(15+4\sqrt{15})$, given by $g_\pm=\epsilon(18+N\pm\sqrt{\Delta})/[2N(N+4)]$, where $\Delta=N^2-180N-540$. 
We are interested in analyzing the quantum critical behavior near the line $f=f_*$. As we have seen, this corresponds to a 
regime of very strong bare gauge coupling. 
%
%\begin{figure}
%[h]
%\centering
% \includegraphics[width=8cm]{vapol.pdf}
% \caption{Feynman diagrams contributing to the vacuum polarization. The solid internal lines represent 
% the propagator $G_0(p)=1/(p^2+m^2)$ and the wiggled lines represent the gauge fields.}
% \label{Fig:vacpol}
%\end{figure}
%
The behavior near the line $f=f_*$ should correspond to a crossover to the critical behavior of the CP$^{N-1}$ model (\ref{Eq:CP1(N-1)}). In order 
to understand this quantum critical 
behavior, we recall that generally near a second order phase transition $m\sim(g-g_*)^{1/\omega}$, where $g_*$ is the infrared 
stable fixed point and $\omega$ is the exponent governing corrections to scaling \cite{ZJ}. In our case, $g_*=g_+$ 
for $N>N_c$ and $\omega=\partial \beta_g(g_+,f_*)/\partial g=\epsilon\sqrt{\Delta}/N$. Due to the presence of the tricritical 
point, we must have $g\to g_-$ for $m\gg\Lambda$ in addition to the usual behavior $g\to g_+$ for $m\ll \Lambda$. Thus, the solution 
of Eq.(\ref{Eq:beta-g}) along the line $f=f_*$ has the general form $m/\Lambda=F(g)/F(g_\Lambda)$, where 
$F(x)=[(g_+-x)/(x-g_-)]^{1/\omega}$ and  $g_\Lambda=g|_{m=\Lambda}$. For $N<N_c$ we have that $\beta_g(g,f_*)\neq 0$ for all 
$g\in\mathbb{R}$, since the fixed points $g_+$ and $g_-$ both become complex. On the other hand,   $\partial\beta_g(g,f_*)/\partial g$ 
vanishes for $g=g_c=(g_++g_-)/2={\rm Re}(g_+)$. Since for $N < N_c$  $g_\pm$ are complex conjugate to each other, $m$ does not exhibit 
a power-law behavior any longer. Indeed, we obtain 
\begin{equation}
 \label{Eq:gap-2}
 F(g)=\exp\left\{-\frac{N}{\epsilon\sqrt{|\Delta|}}\arctan\left[\frac{\sqrt{|\Delta|}\epsilon}{2N(N+4)(g-g_c)}
 \right]\right\}.
\end{equation}
The limit $\epsilon\to 0$ corresponds to a Gaussian fixed point. In the limit $\epsilon\to 0$ we have  $F(g)=\exp\{-1/[(N+4)g]\}$ for all 
$N$. The CP$^{N-1}$ model has a similar behavior at its critical dimension, $d=1+1$.  

We note that $m$ does not vanish at $g=g_c$. As $g\to g_c+$ it approaches its minimum value, $m_{\rm min}$,  and jumps abruptly to 
its maximum value, $m_{\rm max}$,  which is attained as $g\to g_c-$. The difference $m_{\rm max}-m_{\rm min}$ is much larger than $m_{\rm min}$, 
showing that $m$ almost vanishes as $g_c$ is approached from above. Thus, adhering to the logic of the $\epsilon$-expansion, we can 
write approximately,
\begin{equation}
 \label{Eq:gap-3}
 F(g)\approx\exp\left[-\frac{1}{2(N+4)(g-g_c)}\right],
\end{equation}
which vanishes as $g\to g_c+$. On the other hand, approaching $g_c$ from below causes $m$ to grow to infinity.  
This is precisely the type of behavior arising in theories undergoing a CPT \cite{ConfPhTrans}, associated with the 
breakdown of conformal symmetry. This aspect of gauge theories can be related to the so called trace anomaly \cite{Collins-1977} 
of the stress tensor.

In order to find further signatures of a CPT, we search for universal behavior in physical quantities. The spin stiffness 
$\rho_s$ is a crucial physical observable in DQC. In the case of a CPT, it must have a behavior similar to what is found in 
a BKT transition, where the superfluid stiffness exhibits a universal jump at the critical point \cite{NelKost}.  

\begin{figure}
\centering
%\begin{center}
\includegraphics[width=8.5cm]{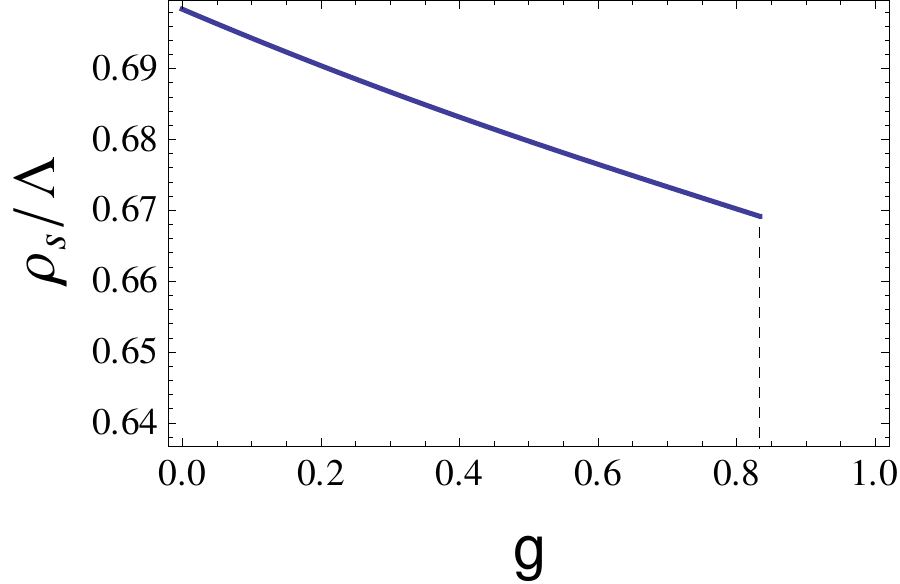}
%\end{center} 
\caption{Spin stiffness obtained by inserting Eq. (\ref{Eq:gap-2}) into  Eq. (\ref{Eq:rhos-1st}) and setting $N=2$ and $\epsilon=1$. There is a universal 
jump at $g=g_c$.}
\label{Fig:spin-stiff}
\end{figure}

To facilitate computing $\rho_s$ within the present formalism, we observe that in the Higgs phase the renormalized photon mass is given by $m_A^2=2e_R^2\rho_s$ 
and use the fact that $m^2/m_A^2=g/(2f)$ to derive an  RG equation 
for $\rho_s$, 
%
%\begin{equation}
%\label{Eq:rhos}
 $md\rho_s/dm=(2-\epsilon-\beta_g/g)\rho_s$, 
%\end{equation}
and solve it over the line $f=f_*$. The solutions have the scaling form, $\rho_s=m^{2-\epsilon}R(m/\Lambda)$.  
Consider first the case having $N>N_c$, where a second-order phase transition takes place. We obtain the typical 
Josephson scaling, including corrections to scaling behavior
\begin{equation}
\label{Eq:rhos-2nd}
 \frac{\rho_s}{\Lambda^{2-\epsilon}}=\frac{[F(g)/F(g_\Lambda)]^{2-\epsilon}\{1+[F(g)]^\omega\}}{g_++g_-[F(g)]^\omega}.
\end{equation}
When $N<N_c$, on the other hand, we have
\begin{equation}
\label{Eq:rhos-1st}
\frac{\rho_s}{\Lambda^{2-\epsilon}}=\frac{2(N+4)[F(g)/F(g_\Lambda)]^{2-\epsilon}\cos\theta(g)}{\sqrt{(N+4)^2g_c^2+\omega^2}\cos[\theta(g)+\theta_0]},
\end{equation}
where $\theta(g)=(1/2)\arctan\{|\omega|/[2(N+4)(g-g_c)]\}$ and $\theta_0=\arctan\{|\omega|/[(N+4)g_c]\}$. 
Since now $g$ must approach $g_c$ from below, it is not possible to use Eq. (\ref{Eq:gap-3}) in Eq. (\ref{Eq:rhos-1st}).  
As a consequence, as $g\to g_c-$ a universal jump arises, which is given by 
%
%\begin{equation}
$\rho_s^c /\Lambda^{2-\epsilon}=2(N+4)[F(g_c-)/F(g_\Lambda)]^{2-\epsilon}/[(N+4)g_c+|\omega|]$. 
%\end{equation}
Thus, we have obtained another expected feature of a CPT reminiscent of the BKT behavior \cite{NelKost}. 
In Fig. \ref{Fig:spin-stiff} we plot $\rho_s$ for $N=2$ and $\epsilon=1$.   

When expressed in terms of $m$, Eq. (\ref{Eq:rhos-1st}) includes logarithmic corrections  to scaling, a behavior related to MC simulations of the 
$J-Q$ model \cite{Sandvik-2010,Damle} and discussed recently in a large $N$ context in Ref. \cite{Nogueira-Sudbo-2012}. Indeed, recalling that 
$m/\Lambda=F(g)/F(g_\Lambda)$, we can write $\cos\theta(g)=\cos\{(|\omega|/2)\ln[F(g_\Lambda) m/\Lambda]\}$ and a similar expression for 
$\cos[\theta(g)+\theta_0]$. In order to see a possible connection with available numerical results, we may consider a finite-size scaling 
approach within the $\epsilon$-expansion framework \cite{ZJ}. Formally, an analysis would make use of an Abelian Higgs model in a periodic hypercube 
along with the results from the RG analysis in the continuum \cite{ZJ}. However, we can already predict the outcome of such a finite-size scaling 
analysis with the results obtained here. To this end, we consider  a correlation length $\xi=m^{-1}=L$, where $L$ represents the finite size of the 
system. For a finite size $L$ and to lowest order in $\epsilon$, we obtain,
\begin{eqnarray}
 \rho_s & \approx &\frac{2(N+4)L^{\epsilon-2}}{\cos\theta_0\sqrt{(18+N)^2+|\Delta|^2/N^2}}
 \nonumber\\
 &\times &\left\{\frac{1}{\epsilon}+\frac{\sqrt{|\Delta|}}{2N}\tan\theta_0\ln\left[\frac{\Lambda L}{F(g_\Lambda)}\right]\right\}, 
\end{eqnarray}
which for $\Lambda L\gg 1$ behaves like $\rho_s\sim L^{\epsilon-2}\ln(\Lambda L)$, similarly to Ref. \cite{Sandvik-2010} when $\epsilon=1$, 
corresponding to $2+1$ dimensions. Thus, such an observed behavior in numerics may be a sign that for considerably larger system sizes a jump 
arises in the spin stiffness. It is worth emphasizing that such a logarithmic correction is tiny at large $N$. To see this, we consider the 
explicit expression for the amplitude of $ L^{\epsilon-2}\ln(\Lambda L)$ for large $L$,
\begin{equation}
 \rho_s\sim \frac{(N+4)\sqrt{|\Delta|}}{N^2(N+18)^2}L^{\epsilon-2}\ln(L/L_0),
\end{equation}
where $L_0=F(g_\Lambda)/\Lambda$. For $N\to\infty$, the coefficient in the expression above behaves like $\sim 1/N^2$, 
being therefore strongly suppressed. 
This is the reason why a large $N$ approach cannot easily predict a logarithmic correction in the spin stiffness. For $N=2$ corresponding 
to $SU(2)$ quantum antiferromagnets, we obtain $\rho_sL^{2-\epsilon}/\ln(L/L_0)\approx 0.1123$. This value is 
only 8-10 \% smaller than the value of $\rho_s L/\ln(L/L_0)$ at critical point $(J/Q)_c$ of the 
$J-Q$ model calculated numerically in Ref. \cite{Sandvik-2010}.

Another interesting quantity is the N\'eel magnetic susceptibility, $\chi_N(x)=\langle{\bf n}(x)\cdot{\bf n}(0)\rangle$, which 
in terms of spinon fields is given by 
$\chi_N(x)=2[\langle{\bf z}^*(x)\cdot{\bf z}(0){\bf z}(x)\cdot{\bf z}(0)\rangle-N^{-1}\langle |{\bf z}(x)|^2|{\bf z}(0)|^2\rangle]$. 
In order to calculate this quantity, the renormalization of composite operators have to be taken into account, so that two 
renormalization constants are needed, namely, the wavefunction renormalization $Z_z$ for the spinon field and $Z_2'$, accounting 
for the renormalization of the composite operator $z_\alpha^*(x)z_\beta(x)$. Thus, at the critical point we have 
$\chi_N(x)=Z_z/(Z_2'|x|^{4-2\epsilon})$ \cite{Nogueira-2008}. In this case it is useful to define the RG function 
$\gamma_2'=m\partial\ln(Z_2'/Z_z)/\partial m$, which is given at one-loop order by $\gamma_2'=3f-g$ \cite{Nogueira-2008}. 
The corresponding finite-size scaling susceptibility thus satisfies $Ld\ln\chi_N/dL=2(2-\epsilon-\gamma_2')$. In general 
we then have $\chi_N=(L_0/L)^{2-\epsilon+\eta_N}X[\ln(L/L_0)]$. For $N>N_c$ a second-order phase transition takes 
place with $X[\ln(L/L_0)]\sim{\rm const}$ and $\eta_N=2-\epsilon+2g_+-6f_*$. For $N<N_c$, on the other hand, we obtain,
\begin{equation}
 \chi_N=\frac{\{\ln[1/F(g_\Lambda)]/\ln(L/L_0)\}^{1/(N+4)}}{[F(g_\Lambda)L/L_0]^{2-\epsilon+\eta_N}},
\end{equation}
where $\eta_N=2-\epsilon+2g_c-6f_*$. Note, however, that in the latter expression positive values of $\eta_N$ arise only for 
$(13+\sqrt{385})/2<N<N_c$. Thus, a better approximation is necessary in order to access more physical values of $N$. 
Anyway, it is interesting to notice that a nontrivial logarithmic dependence arises in this case.

As a final calculation to support a CPT scenario in DQC, we consider the dynamics of instantons inside the VBS 
in the CP$^{N-1}$ model (\ref{Eq:CP1(N-1)}) at fixed dimensionality and large $N$. At leading order a standard calculation 
yields the mass gap, $M$, which due to the large $N$ limit exhibits, as expected, a conventional 
power-law behavior for $\hat g>\hat g_c$, i.e.,  
%
%\begin{equation}
% \label{Eq:gap-CP1}
 $M/\Lambda=(2/\pi)(1-\hat g_c/\hat g)$,
%\end{equation}
where $\hat g_c=2\pi^2/N$. However, here the unconventional behavior arises in the gapped instanton excitations. By computing the vacuum polarization, 
we obtain the correction to the Maxwell term responsible for  the most important contribution to the instanton dynamics. From the strong-coupling 
regime $e^2\to\infty$ for fixed $d=2+1$, we obtain from the first line of Eq. (\ref{Eq:L-Maxwell-1}) with $m$ replaced by $M$, 
\begin{equation}
 \label{Eq:Maxwell}
 {\cal L}_{\rm Maxwell}\approx\frac{N}{48\pi M}(\epsilon_{\mu\nu\lambda}\partial_\nu A_\lambda)^2.
\end{equation}
A Maxwell Lagrangian in three spacetime dimensions supports instantons, provided the $U(1)$ gauge group is compact. This amounts to considering compact 
electrodynamics \cite{Polyakov}, which is equivalent to a field theory for a Coulomb gas of instantons. The instanton action is given by

\begin{eqnarray}
S_{\rm inst}&=&\frac{N}{48M}\sum_{i\neq j}\frac{q_iq_j}{|x_i-x_j|}+\frac{N\Lambda}{24\pi M}\sum_i q_i^2
\nonumber\\
&-&2N\sum_i\rho_{q_i}\ln\left(\frac{M}{\Lambda}\right),
\end{eqnarray}
where $q_i=\pm 1,\pm 2,\dots$ are instanton-charges. The first two terms are the usual contributions originating with compact electrodynamics in $2+1$ 
dimensions \cite{Polyakov}. The last term was computed in Ref. \cite{Murthy-Sachdev}, and describes the core-contribution to the action of {\it non-interacting} 
instantons. We will consider only  the contribution having $q_i=\pm 1$, which yields $\rho_1\approx 0.06$ \cite{Murthy-Sachdev}. Therefore, the corresponding field 
theory for the instantons is given by the following sine-Gordon Lagrangian

\begin{equation}
{\cal L}_{\rm SG}=\frac{1}{2}(\partial_\mu\varphi)^2
-z\cos(2\pi s\varphi),
\end{equation}
where
%\begin{equation}
$s=\sqrt{N\hat g/[48\Lambda(\hat g-\hat g_c)]}$,
%\end{equation}
and
%
%\begin{equation}
$z=\Lambda^3(\hat g-\hat g_c)^{-2N\rho_1}\exp\{-N\hat g/[48(\hat g-\hat g_c)]\}$ 
%\end{equation}
is the fugacity of the Coulomb gas. Thus, within a Debye-H\"uckel (DH) approximation, we find the screening mass gap of the 
instanton gas given by
\begin{equation}
\label{Eq:M}
M^2_{\rm DH}=4\pi^2s^2z=\frac{\Lambda^2\pi^2N\hat g}{12(\hat g-\hat g_c)^{1+2N\rho_1}}\exp\left[-\frac{N\hat g}{48(\hat g-\hat g_c)}\right].
\end{equation}
Eq. (\ref{Eq:M}) gives the mass gap of the confining dual photon  in the VBS phase. This is not a simple power,
featuring in addition an essential singularity at the critical point, providing further indications of a CPT.   

The VBS order parameter is given within the large $N$ framework by $\psi_{\rm VBS}=\langle e^{i2\pi s\varphi}\rangle$, which 
in the DH approximation is easily evaluated by means of a Gaussian integration, $\psi_{\rm VBS}\approx e^{-2\pi^2s^2\langle\varphi^2\rangle}$, 
where $\langle\varphi^2\rangle\approx \Lambda/(2\pi^2)-M_{\rm DH}/(4\pi)$. Thus, we obtain, 
\begin{eqnarray}
 \psi_{\rm VBS}&\approx &\exp\left\{-\frac{N\hat g}{48(\hat g-\hat g_c)}
 +\frac{\pi^2(N\hat g)^{3/2}}{96\sqrt{3}(\hat g-\hat g_c)^{3/2+N\rho_1}}
 \right.\nonumber\\
 &\times &\left.\exp\left[-\frac{N\hat g}{96(\hat g-\hat g_c)}\right]
 \right\},
\end{eqnarray}
which vanishes continuously as $\hat g\to \hat g_c$ from above. In terms of the correlation lengths $\xi=M^{-1}$ and 
$\xi_{\rm mon}=M_{\rm DH}^{-1}$ for the instantons, we obtain  
in the present approximation and for $\hat g$ near 
$\hat g_c$, $\psi_{\rm VBS}\sim \xi^{-(1+2N\rho_1)}\xi_{\rm mon}^{-2}$.  If $\xi_{\rm mon}$ is ignored, it would appear that the 
VBS correlation length $\xi_{\rm VBS}$ has a power-law behavior relative to $\xi$. However, due to $\xi_{\rm mon}$, we find that 
the actual behavior near the critical point is highly peaked and vanishes quickly for $\hat g$ not much larger than $\hat g_c$. This 
result might be an artifact associated to the missing magnitude of $\psi_{\rm VBS}$ in the above approximation.  

To test numerically for a CPT, one needs to establish a universal jump in the stiffness at the transition. This is similar to the situation 
in the $2D$ XY model, which features a BKT transition, which is a CPT. This is done by considering higher-order 
response functions to phase-twists \cite{2DCPT}. Similar techniques have been developed for searching for universal jumps in stiffnesses 
in 3D systems with proposed CPTs \cite{3DCPT}.

Summarizing, 
%to obtain an improved understanding of deconfined quantum criticality, 
we have analyzed the $\epsilon$-expansion of the Abelian Higgs 
model in the allegedly first-order phase transition regime along a line in the RG flow diagram determined by the gauge coupling fixed points 
defining the strong-coupling regime. We have  argued that within the accuracy of the $\epsilon$-expansion, a conformal phase transition associated 
with a deconfined quantum critical point occurs. We obtain a spinon mass gap featuring an essential singularity at the critical point. Similarly 
to the BKT transition in two dimensions, we find that the spin stiffness has a universal jump at the conformal phase transition critical point. 
We find further evidence for a conformal phase transition by analyzing the VBS phase at large $N$ in the presence of instantons, 
where the screening mass of the instantons also exhibits an essential singularity at the critical point.

\acknowledgments

F.S.N.  acknowledges the Deutsche Forschungsgemeinschaft (DFG) for the financial support via the 
collaborative research center SFB TR 12. A.S. acknowledges support 
from the Research Council of Norway, Grant Nos. 205591/V20 and 216700/F20.

\end{document}